\documentclass{ws-procs961x669}            

\begin{document}
\title{Pion nucleus Drell-Yan process and parton transverse momentum in the pion
	}

\author{A.~Courtoy$^*$}

\address{Instituto de F\'isica, Universidad Nacional Aut\'onoma de M\'exico\\
Apartado Postal 20-364, 01000 Ciudad de M\'exico, Mexico\\
$^*$E-mail: aurore@fisica.unam.mx}

\begin{abstract}
During the INT-18-3 workshop, we presented an analysis of unpolarized Drell-Yan
pair production in pion-nucleus scattering with a particular focus into the pion Transverse Momentum Distributions in view of the future Electron Ion Collider. 
The  transverse distributions of
the pion calculated in a Nambu--Jona-Lasinio framework, with Pauli-Villars regularization, were used. 
The pion Transverse Momentum Distributions evolved  up to next-to-leading logarithmic accuracy is then tested against the transverse momentum spectrum
of dilepton pairs up to a transverse momentum of 2 GeV. 
  We found a  fair agreement with available
pion-nucleus data. 
This contribution joins common efforts from the TMD  and the pion structure communities for the Electron Ion Collider.
\end{abstract}


\bodymatter

\section{Introduction}
The structure of the pion is intrisically related to the dynamics of QCD. Non-perturbative approaches that incorporate chiral symmetry have been successfully leading towards a more general understanding of the pion structure.  
While there are less experiments involving pion observables than, {\it e.g.}, proton's, an Electron Ion Collider (EIC) will be capable of  providing more experimental insights, as described in Ref.~\cite{Aguilar:2019teb}. In these proceedings, we will focus on a calculation of the pion Transverse Momentum Distributions (TMDs) in the Nambu--Jona-Lasinio (NJL) model, whose results fully embody chiral symmetry.

The pion-induced Drell-Yan process has played a central role in the early determination of the pion distributions, especially through the data available in Ref.~\cite{Conway:1989fs}.   
This process is described by $h_1(p_1) \; h_2(p_2)\to \gamma^*(q)+X$
in which a virtual photon is produced with large 
invariant mass-squared $Q^2$ in the 
collisions of two hadrons at a center-of-mass energy $s = (p_1 + p_2)^2$, with $p_{1,2}$ the four momentum of hadrons $h_{1,2}$, 
 $h_1$  being a pion and $h_2$ a proton. 
 Beyond collinear approaches, the  unintegrated cross-sections characterize the spectrum of transverse momentum of the virtual photon, $q_T$. In the kinematical regime in which $q_T$ is of order $\Lambda_{\mbox{\tiny QCD}}$, that is  small {\it w.r.t.} $Q$, such an effect is accounted into transverse momentum of the partons through TMDs. The departure from collinearity is here a highly non-perturbative effect, originated in the internal dynamics of the parent hadron.
While there exist model predictions as well as phenomenological analyses, the implementation of the transverse momentum factorization theorems has added in complexity in globally fitting and phenomenologically determining TMDs.

In Ref.~\cite{Ceccopieri:2018nop}, we presented one of the first analyses of the $\pi N$ DY process in terms of the modern TMD formulation. Our method focuses on investigating the DY cross section from the perspective of the dynamics of the pion as expressed in the Nambu--Jona-Lasinio (NJL) model~\cite{ns}.

\section{Pion dynamics in Drell-Yan cross sections}
When $q_T^2$ becomes small compared to $Q^2$, large logarithmic 
corrections of the form of $\alpha_s^n \log^m(Q^2/q_T^2)$ with $0 \leq m \leq 2n-1$
appear in fixed order 
results, being $n$ the order of the perturbative calculation. 
These large logarithmic corrections can be resummed to all
orders by using the Collins-Soper-Sterman (CSS) formalism~\cite{CSS}.
In this limit, the 
cross-section, differential in ${\bf q}_T$, can be written as~\cite{Ceccopieri:2014qha} 
\begin{eqnarray}
\label{CSS}
\frac{d\sigma}{dq_T^2 d\tau dy}&=& \sum_{a,b} \sigma_{q\bar{q}}^{(LO)}  
\int_0^\infty db \frac{b}{2} J_0(b \, q_T)\,S_q(Q,b)\, S_{NP}^{\pi p}(b)   \nonumber\\
&& \Big[ (f_{a/\pi}\otimes C_{qa})\left(x_1,\frac{b_0^2}{b^2}\right) 
(f_{b/p}\otimes C_{\bar{q}b})\left(x_2,\frac{b_0^2}{b^2}\right)+ q \leftrightarrow  \bar q \Big]\,,
\end{eqnarray}
where $b_0=2 e^{-\gamma_e}$, the symbol $\otimes$ stands for convolution 
and $\sigma_{q\bar{q}}^{(LO)}$ is the leading-order 
total partonic cross section for producing a lepton pair.
We use the nuclear PDFs 
of Ref.~\cite{nCTEQ} for the proton collinear PDFs. 
The cross section in Eq.~(\ref{CSS}) is also differential in $\tau=Q^2/s$ and $y$, the rapidity of the DY pair. 
The large logarithmic corrections are exponentiated 
in $b$-space in the Sudakov perturbative form factor, $S_q(Q,b)$.
The non-perturbative factor, $S_{NP}^{\pi p}$, contains all the information about the non-perturbative ${\bf k}_T$ behavior. Models for hadron's structure might describe the true theory but at one specific value of the RGE scale, $Q_0$. In that sense, they incorporate ---here, in the chiral limit---  a factorized ${\bf k}_T$  and Bjorken $x$ behavior in a fashion that can be resumed as 
\begin{eqnarray}
f^{q/\pi}(x_{\pi}, b; Q_0^2)&=& q(x_{\pi}; Q_0^2) \,S_{NP}^{\pi}(b)\,=\,q(x_{\pi}; Q_0^2)  \exp\{\ln S_{NP}^{\pi}(b)\}~;
\label{chi_sudakov}
\end{eqnarray}
while the full  TMD parton densities should involve a further $Q^2$ dependence, that is also called   {\it non-perturbative} evolution. Such an evolution has been parameterized in the past, for the proton-proton DY, as
\begin{eqnarray}
\label{fnp}
S_{NP}^{pp}(b)
&=&\exp\{-[a_1 + a_2 \ln (M/(3.2 \,\mbox{GeV})) + a_3 \ln(100 x_1 x_2)] b^2\}\,,
\end{eqnarray}
where $a_1$ plays an equivalent r\^ole to $\ln S_{NP}^{\pi}(b)$ in Eq.~(\ref{chi_sudakov}) and the other parameters, determined  {\it e.g.} in Ref.~\cite{KN05},  reflect a $Q^2$ evolution, {\it i.e.} through the invariant mass $M$ dependence, as well as an unfactorized $(x, {\bf k}_T)$ term.

Our ansatz, incorporating the NJL results, is 
\begin{equation}
\label{prescript_us}
S_{NP}^{\pi p}(b)=S_{NP}^{\pi}(b) \, \sqrt{S_{NP}^{pp}(b)} \,,
\end{equation}
where $S_{NP}^{\pi}(b)$ is given by the model and is non-gaussian, {\it i.e.}
\begin{eqnarray}
S_{NP}^{\pi}(b)
&=& \frac{3}{2 \pi^2} {\left( m \over f_\pi  \right )^2}~
\sum_{i=0,2} c_i K_0 ( m_i \, b)\,,
\label{bsp}  
\end{eqnarray}
and the square root of $S_{NP}^{pp}(b)$ is  given 
in Eq.~(\ref{fnp}).
 
 At very large values of $b$, the perturbative form factor needs to include a {\it taming} through the so-called $b^{\star}$-prescription.
By  splitting the perturbative
form factor, we can use distinct $b_{max}$ on 
the proton and pion side with $b_\star(b,b_{max})=b/\sqrt{1+\Big(\frac{b}{b_{max}}\Big)^2} $ and the respective $b_{max}^p=1.5$GeV$^{-1}$ and  $b_{max}^{\pi}=b_0/Q_0=2.44$GeV$^{-1}$, with $Q_0$ ---the hadronic scale of the model--- evaluated at NLO through a minimization procedure for the collinear PDF on the integrated DY cross-sections. That hadronic scale so determined represents the only free parameter of our approach.

\begin{figure}[h]
\begin{center}
\includegraphics[width=5in]{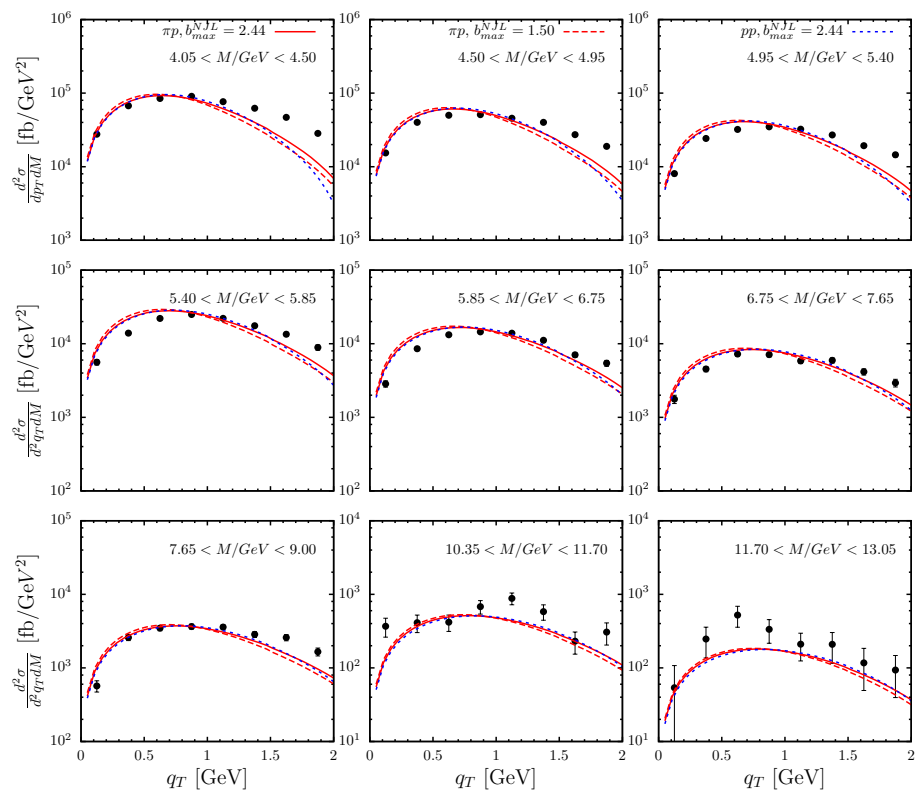}
\end{center}
\caption{Our model results compared to cross sections
in various invariant mass bins of the DY lepton pair  integrated in $0<x_F<1$. $b_{max}$ is given in unit GeV$^{-1}$.}
\label{Fig:DYqTNJL}
\end{figure}

 The results  for the lepton pair $q_T$-spectra of Ref.~\cite{Conway:1989fs}, measured in $\pi W$ collisions, are shown in Fig.~\ref{Fig:DYqTNJL}. The plots range up to $q_T \sim 2$ GeV, range for which the chosen proton description holds~\cite{Ceccopieri:2018nop}. Both red, full and dashed, curves correspond to results using Eq.~(\ref{prescript_us}) with, respectively, the proposed regulator value $b_{max}^{\pi}=2.44$ GeV$^{-1}$ and $b_{max}^{\pi}=1.5$ GeV$^{-1}$ demonstrating the stability of our results at small values of $q_T$. The short-dashed blue curve corresponds to a different ansatz: Eq.~(\ref{fnp}) is used for both hadrons, still with different $b_{max}$ values. At low-$q_T$, the difference between the two ans\"atze is quantitatively small. There seems to be a reduced sensitivity of the data 
to non-perturbative structure. However, this is a first analysis for which no {\it non-perturbative evolution} of the type  Eq.~(\ref{fnp}) has been included, since it is not inherent to the NJL approach used here.

\section{Conclusions}

We have synthetized the analysis of the DY pair production
in pion-nucleus scattering~\cite{Ceccopieri:2018nop}, in which we tested the outcome of a NJL approach for the pion TMD plugged in the CSS framework  at next-to-leading logarithmic accuracy against the differential transverse momentum
spectra of DY pairs produced in $pA$ collisions. 
Further analyses require an extension of the current approach to include a $Q^2$ dependence. This will be particularly relevant for both the pion structure and the TMD formalism that will be addressed at the EIC.

\section*{Acknowledgments}

The author thanks her colleagues and co-authors of the original publication. This work has been funded by UNAM through the PIIF project Perspectivas en F\'isica de Part\'iculas y Astropart\'iculas as well as  Grant No. DGAPA-PAPIIT IA102418.

\bibliographystyle{ws-procs961x669}

\end{document}